\documentclass[12pt]{article}
\usepackage{amsmath,amssymb,amsthm,color}
\usepackage{graphicx}
\usepackage{cite}
\usepackage{epsfig}
\usepackage{lineno}
\renewcommand{\baselinestretch}{1.66}

\begin{document}

\title {Unique electronic and optical properties of stacking-modulated bilayer graphene under external magnetic fields}

\author{
\small Chiun-Yan Lin$^{a}$, Da-We Weng$^{b}$, Chih-Wei Chiu$^{b}$, Godfrey Gumbs$^{*c}$\\
\small $^{a}$Department of Physics, National Cheng Kung University, Taiwan\\
\small $^{b}$Department of Physics, National Kaohsiung Normal University, Kaohsiung, Taiwan\\
\small $^{c}$Department of Physics and Astronomy, Hunter College of the City University of New York,\\
\small Park Avenue, New York, New York, USA\\
}
\renewcommand{\baselinestretch}{1.66}
\maketitle

\renewcommand{\baselinestretch}{1.66}

\begin{abstract}

This study delves into the magneto-electronic and magneto-optical properties of stacking-modulated bilayer graphene. By manipulating domain walls (DWs) across AB-BA domains periodically, we unveil oscillatory Landau subbands and the associated optical excitations.  The DWs act as periodic potentials, yielding fascinating 1D spectral features. Our exploration reveals 1D phenomena localized to Bernal stacking, DW regions, and stacking boundaries, highlighting the intriguing formation of Landau state quantization influenced by the commensuration between the magnetic length and the system. The stable quantized localization within different regions leads to the emergence of unconventional quantized subbands.  This study provides valuable insights into the essential properties of stacking-modulated bilayer graphene.

\end{abstract}
\par\noindent ~~~~$^*$Corresponding author- E-mail: ggumbs@hunter.cuny.edu (G. Gumbs)

\pagebreak
\renewcommand{\baselinestretch}{2}
\newpage

\vskip 0.6 truecm

\section{Introduction}

Bilayer graphene (BLG) has garnered substantial attention due to its intriguing physical and optical characteristics \cite{RPP76;056503,PhysRep648;1}. Among the factors that significantly impact these essential properties under external influences, the stacking configuration stands out as a key player.
Recent advancements in the precise control of few-layer van der Waals 2D materials have brought about extensive investigations into the effects of relative displacement between two graphene sheets, particularly in the subject of twistronics\cite{NatMater21;844,PRL108;205503} or straintronics\cite{npjQMater6;59}. The phenomenon of structural superlubricity \cite{Nature563;485} allows for interlayer sliding, resulting in complex stacking patterns, particularly achieved in graphene flakes grown through chemical vapor deposition (CVD) methods.
This manipulation has led to various categories of BLG systems, such as twisted \cite{AdvMater33;2004974}, sliding \cite{ACSNANO7;1718}, and geometry-modulated systems through domain boundaries \cite{NatComm7;11760,NatNanotechnol11;1060,NatNanotechnol13;204,PNAS110;11256,NanoLett13;3262}. Such engineered stacking symmetries can dramatically alter electronic behavior and interband excitations, even with slight misalignments between adjacent layers.

Two particularly inspiring examples, sliding and twisted BLG, have been synthesized with precise control of stacking configurations, each showcasing 2D phenomena \cite{Nature556;43,PRB87;205404,SciRep4;7509,PRB84;155410,SciRep5;17490,PRB104;205121}. These systems exhibit unique physical properties that set them apart from normal bilayer systems. Sliding BLG, for instance, presents hybridized Dirac cones, specific Landau levels(LLs)\cite{SciRep4;7509} and optical properties \cite{SciRep4;7509,PRB104;205121}. It also demonstrates electronic topological transitions under an electric field \cite{SciRep5;17490,PRB84;155410}. On the other hand, twisted BLG, forms a Moire superlattice with intriguing 2D energy subbands \cite{PRB87;205404}, giving rise to remarkable phenomena like Moire superconductivity \cite{Nature556;43} and undamped interband plasmons \cite{Nature605;63,NanoLett16;6844}. The inherent stacking symmetries in these systems unveil intriguing and potentially useful properties that have captivated researchers worldwide.

BLG predominantly exhibits two prototype stacking configurations, hexagonal (AA) and Bernal (AB), due to weak interlayer interactions \cite{RPP76;056503}. Large CVD-grown areas primarily adopt Bernal stacking with embedded layer-stacking domain walls (DWs) that transition between AB and BA configurations\cite{PNAS110;11256,NanoLett13;3262,NatNanotechnol13;204}.
Owing to the varied local interlayer potentials across the Bernal/DW boundary, the stacking-modulated crystalline structure can be observed by scanning tunneling microscopy (STM)\cite{NatComm7;11760}, scanning transmission electron microscopy (STEM)\cite{PNAS110;11256,NanoLett13;3262} or infrared nanoscopy\cite{NatNanotechnol13;204}.
These DWs introduce topological edge states as a result of 1D DW solitons in BLG\cite{PRX3;021018,NatNanotechnol13;204}.
Experimental investigations have confirmed intriguing phenomena such as topological valley transport \cite{NanoLett20;5936}, topologically protected valley Hall edge states \cite{PRB105;L081408}, and plasmon reflections by topological boundaries\cite{NanoLett17;7080}. Additionally, precise control of stacking boundaries has been achieved through systematic geometric modulation using an atomic force microscope tip\cite{NatNanotechnol13;204}. This controlled transformation from AB to BA stacking, mediated by DWs typically ranging from several to dozens of nanometers, presents exciting prospects for electronic applications\cite{NatNanotechnol13;204}.

Scanning tunneling spectroscopy (STS) has provided experimental confirmation of the presence of the 1D edge states in BLG under the influence of magnetic fields\cite{NatComm7;11760}. Additionally, band inversions occurring at DWs within gated BLG systems result in protected crossings of valence and conduction bands near the Fermi level ($E_{F}=0$), giving rise to topological states with distinctive magnetic confinements observed in AB, BA, and DW regions\cite{PRX3;021018}. Theoretical investigations employing first-principles methods\cite{PRX3;021018} and the tight-binding model\cite{SciRep9;859} have been instrumental in examining the energy bands of AB/DW/BA/DW BLG. Previous research\cite{SciRep9;859} has discussed the profound influence of stacking modulation on BLG's electronic properties. Stacking modulation introduces 1D parabolic dispersions, van Hove singularities (vHSs), and distinct absorption peaks that are dependent on the DW width. The creation of 1D topological states leads to semi-metallic behavior and a blue shift in absorption peaks, indicative of topological band crossings. These findings present exciting opportunities to explore and comprehend the unique properties of stacking-modulated BLG under the influence of external magnetic fields.

In this study, we employed the tight-binding model and Kubo formula to explore the magnetic effects on the electronic properties and optical excitations in the geometry-modulated BLG. The main purpose of this work is to explore the interaction between magnetic field effects and stacking DWs. We focus on unveiling stacking-modulated properties in BLG, with a specific emphasis on the length scale between the magnetic length ($l_{B}$) and the system. The generalized tight-binding model is used to address the non-uniform magnetization induced by stacking modulation, allowing the inclusion of non-uniform hopping integrals and Peierls phases in the calculations\cite{ACSNano4;1465,CARBON69;151}.
Oscillatory Landau subbands (LSs) and extra optical excitations are revealed. The intriguing nature of magnetically quantized states emerges, with the influence of DWs giving rise to different features associated with the local stacking symmetries. The interaction range is affected by the ratio between $l_{B}$ and the system scale of the Bernal and DW size. Stacking boundaries play an important role in magnetic and quantum confinement effects. The theoretical predictions can be experimentally examined using STM and STS\cite{NatComm7;11760,NatPhys3;623} and magneto-optical spectroscopies\cite{PRL102;166401,PRL100;087403}. These investigations are essential to shed new light on the unique physical and optical properties of BLG under external influences.
Comprehensive comparisons of magneto-optical properties under different field strengths and DW/stacking configurations enable us to tailor the material's properties for a wide range of applications, including quantum devices and optoelectronic technologies.

\section{Generalized tight-binding model: magneto-electronic and magneto-optical calculations}


The non-uniform magnetic quantization caused by the geometric modulation of DWs is thoroughly explored in BLG. A schematic AB/DW/BA/DW modulation is depicted in Fig. 1(a) from a top view. The DWs are formed along the armchair direction ($\widehat{x}$) through a slow and uniform variation of the C-C bond length $b=1.42$ ${\AA}$ to cause stacking transformation from AB to BA. The DWs are assumed to be placed periodically, and the two uniform Bernal regions (AB or BA) beside the DW are of equal length. This system is characterized by parameters ($N_{B}$, $N_{D}$), which indicate the length of the desired regions with width of $3N_{B}b$ and $3N_{D}b$, respectively. The carbon atoms in a primitive unit cell are denoted as $A^{j}_{i}$ and $B^{j}_{i}$, where the lattice index $i$ ranges from 1 to $4(N_{B}+N_{D})$ and the layer index $j$=1 and 2. In the Bernal stackings, $A^{1}_{i}$ ($B^{1}_{i}$) is located on top of $A^{2}_{i}$ ($B^{2}_{i}$) for AB (BA) stacking. Random stackings cause the destruction of hexagonal symmetries along $\widehat{x}$ but maintain translation symmetry along the $\widehat{y}$.


The generalized tight-binding model has been developed to comprehend the electronic properties of BLG. In the absence of external fields, the Hamiltonian is given by the following expression:
\begin{equation}
H=-\sum_{{\bf R_{i}},{\bf R_{i}}}t({\bf R_{i}},{\bf R_{j}})[c^{+}({\bf R_{i}})c({\bf R_{j}})+h.c.],
\end{equation}
where $c^{+}({\bf R_{i}})$ and $c({\bf R_{j}})$ are the creation and annihilation operators at lattices ${\bf R_{i}}$ and ${\bf R_{j}}$, respectively. The hopping integrals $t$ can be calculated using the empirical formula:
\begin{equation}
-t({\bf R_{i}},{\bf R_{j}})=\gamma_0 e^{-\frac{d-b}{\rho}}(1-(\frac{\mathbf{d}\cdot \widehat{z}}{d})^2) + \gamma_1 e^{-\frac{d-d_0}{\rho}} (\frac{\mathbf{d}\cdot \widehat{z}}{d})^2,
\end{equation}
where $\mathbf{d}$ is the vector connecting two lattice points ${\bf R_{j}}-{\bf R_{i}}$, $d_0=3.35$ {\AA} is the interlayer distance, and $\rho = 0.184 b$ is the characteristic decay length. $\gamma_0 = -2.7$ eV and $\gamma_1 = 0.48$ eV represent the hoppings for nearest-neighbor intralayer and vertical interlayer atoms, respectively\cite{SciRep4;7509,SciRep9;859,PRB87;205404}. In this study, the cutoff for transfer integrals is implemented within a circle projected onto the graphene plane with a radius of $2b$. Consequently, the remote integrals, which are two orders of magnitude smaller, can be safely neglected\cite{SciRep4;7509,SciRep9;859}.

The wave function in the stacking-modulated BLG is represented as a linear combination of tight-binding functions:
\begin{equation}
\Psi = C_{A^{j}_{i}} \varphi_{A^{j}_{i}(\mathbf{k})} + C_{B^{j}_{i}} \varphi_{B^{j}_{i}(\mathbf{k})} ,
\end{equation}
where $ \varphi_{A^{j}_{i}(\mathbf{k})} = \sum^{4(N_{AB}+N_{D})}_{i=1} \sum^{2}_{j=1}\exp (i\mathbf{k}\cdot \mathbf{R}_{A^{j}_{i}})
\chi (\mathbf{r}-\mathbf{R}_{A^{j}_{i}})$ and $ \varphi_{B^{j}_{i}(\mathbf{k})} = \sum^{4(N_{AB}+N_{D})}_{i=1}\sum^{2}_{j=1} \exp (i\mathbf{k}\cdot \mathbf{R}_{B^{j}_{i}})
\chi (\mathbf{r}-\mathbf{R}_{B^{j}_{i}})$ represent the tight-binding functions constructed from 2$p_{z}$ atomic orbitals $\chi$ of the lattice sites.
In the tight-binding scheme, the Hamiltonian can be expressed as a ${16(N_{AB}+N_{D})\times16(N_{AB}+N_{D})}$ Hermitian matrix and separated into  elements from intralayer hoppings and interlayer hoppings, defined as follows:
\begin{equation}
h_{A^{j}_{i}A^{j}_{i^{\prime}}}(\mathbf{k})=\langle \varphi _{A^{j}_{i}(\mathbf{k})}|H|\varphi _{A^{j}_{i^{\prime}}(\mathbf{k})}\rangle  \\
\end{equation}
\begin{equation}
h_{A^{j}_{i}B^{j}_{i^{\prime}}}(\mathbf{k})=\langle \varphi _{A^{j}_{i}(\mathbf{k})}|H|\varphi _{B^{j}_{i^{\prime}}(\mathbf{k})}\rangle  \\
\end{equation}
\begin{equation}
h_{A^{1}_{i}B^{2}_{i^{\prime}}}(\mathbf{k})=\langle \varphi _{A^{1}_{i}(\mathbf{k})}|H|\varphi _{B^{2}_{i^{\prime}}(\mathbf{k})}\rangle  ,
\end{equation}
where $A$ and $B$ sublattices are alternating. The on-site energies and the overlap integrals are assumed to be zero. When $\mathbf{d}$ is perpendicular to $\widehat{z}$, only intralayer couplings, as in Eq. (4) and Eq. (5), persist, with consideration limited to $d_{c}\leq 2b$ for both the Bernal and DW regions. In this scenario, the nearest neighboring term occurs in Eq. (5) when $|\mathbf{d}|=b$, expressed with a phase $e^{-ik_{x}(b+\delta)}$ for $i=i^{\prime}$, and $2e^{ik_{x}(b/2+\delta)}\cos(\frac{\sqrt{3}k_{y}b}{2})$ for $i^{\prime}=i+1$. Here, $\delta=0$ for the Bernal region and $\delta=b/(4N_{D}-1)$ for the DW region, respectively. For longer distance hoppings, e.g., the next-nearest neighboring intralayer terms, they follow a similar expression for the phase, which can be easily calculated as they decay by the factor $e^{-\frac{d-b}{\rho}}$. However, the interlayer Hamiltonian elements in Eq. (6) display more complex, involving interaction cutoffs within $d_{c}$. To compute all the effective hopping energies and the sum of all phase terms, numerical simulation methods are employed.

When considering the growth of real materials, it is reasonable to assume that the Bernal width is greater than DW width. In this study, we configure a system consisting of approximately a hundred cells of each Bernal domain and DWs with sizes ranging from ~5 $nm$ to ~30 $nm$. The geometry-modulated BLG exhibits a quasi-1D energy band structure defined within the first Brillouin zone, where $-\pi/a\leq k_{y} \leq \pi/a$, and $a=\sqrt{3}b$ represents the lattice constant of a monolayer graphene.

In the presence of a magnetic field $\mathbf{B}=B\widehat{z}$, using the Landau gauge $\mathbf{A}=(0,Bx,0)$, the ${\bf B}$-induced Peierls phase is given as follows\cite{PhysicaE40;1722}:
\begin{equation}
G(\mathbf{R}) \equiv\frac{2\pi}{\phi_0} \int_{\mathbf{R^{\prime}}}^{\mathbf{R}} \mathbf{A(r^{\prime})}\cdot d\mathbf{r^{\prime}} ,
\end{equation}
where $\phi_0 = hc/e$ represents the magnetic flux quantum. The phase is defined across both the uniform AB (BA) region and the non-uniform DW region. We focus on the scenarios where the length scale of the DW is commensurate with the magnetic length $l_{B}$, which requires the AB/DW cell to be enlarged by a factor of $N_{sc}$ to accommodate the $\phi_0$ flux\cite{PhysicaE40;1722}. The parameter $N_{sc}$ characterizes the periodic length of the modulated configuration, specifically causing the period $2\pi/N_{sc}a$ in the energy spectrum.

Equations 4-6, which incorporate the Peierls phase, are central to our analysis:
\begin{equation}
h_{A^{j}_{i}A^{j}_{i^{\prime}}}(\mathbf{k})=\exp i[G(R_{A^{j}_{i^{\prime}}})-G(R_{A^{j}_{i}})] \langle \varphi _{A^{j}_{i}(\mathbf{k})}|H|\varphi _{A^{j}_{i^{\prime}}(\mathbf{k})}\rangle  \\
\end{equation}
\begin{equation}
h_{A^{j}_{i}B^{j}_{i^{\prime}}}(\mathbf{k})=\exp i[G(R_{B^{j}_{i^{\prime}}})-G(R_{A^{j}_{i}})] \langle \varphi _{A^{j}_{i}(\mathbf{k})}|H|\varphi _{B^{j}_{i^{\prime}}(\mathbf{k})}\rangle  \\
\end{equation}
\begin{equation}
h_{A^{1}_{i}B^{2}_{i^{\prime}}}(\mathbf{k})= \exp i[G(R_{B^{2}_{i^{\prime}}})-G(R_{A^{1}_{i}})] \langle \varphi _{A^{1}_{i}(\mathbf{k})}|H|\varphi _{B^{2}_{i^{\prime}}(\mathbf{k})}\rangle .
\end{equation}
The variation of phase terms are relevant when considering the bond length distortion $\delta=b/(4N_{D}-1)$. In this approach, the energy bands and wave functions are obtained, ultimately allowing us to calculate the optical spectral function.

The optical spectral function, which accounts for the absorption of a photon and the excitation of electronic states, can be derived using the Fermi's golden rule. It can be expressed as follows:
\begin{eqnarray}
A(\omega )&\propto&\sum_{n^{v},n^{c}}^{}\int_{1stBZ}\frac{d\textbf{k}}{{
2\pi }*{2\pi }}
\left\vert \left\langle \Psi ^{c}_{\textbf{k}}(n^{c})\left\vert \frac{
\widehat{\mathbf{E}}\cdot \mathbf{P}}{m_{e}}\right\vert \Psi ^{v}_{\textbf{k}}(n^{v})\right\rangle \right\vert^{2} \\ \nonumber%
&\times& Im\left\{\frac{{
f[E^{c}_{\textbf{k}}(n^{c})]-f[E^{v}_{\textbf{k}}(n^{v})]}} {E^{c}_{\textbf{k}}(n^{c})-E^{v}_{\textbf{k}}(n^{v}){-\omega }{-\imath \Gamma }} \right   \} ,
\end{eqnarray}
where $\textbf{E}$ represents electric polarization, $\textbf{P}$ denotes momentum operator, $f[E_{k_{y}}(n)]$ corresponds to the Fermi-Dirac distribution, $m_{e}$ stands for the electron mass and $\Gamma$ is the phenomenological broadening parameter. The above equation involves the joint density of states (DOS) and dipole transition from $E^{v}_{\textbf{k}}(n^{v})$ to $E^{c}_{\textbf{k}}(n^{c})$. The former is well described by the tight-binding model. On the other hand, the latter is calculated using the gradient approximation, expressed as $\left\langle \Psi ^{c}_{\textbf{k}}(n^{c})\left\vert \frac{
\widehat{\mathbf{E}}\cdot \mathbf{P}}{m_{e}}\right\vert \Psi ^{v}_{\textbf{k}}(n^{v})\right\rangle\simeq \frac{\partial}{\partial k_{y}}\left\langle \Psi ^{c}_{\textbf{k}}(n^{c})\left\vert \mathbf{H} \right\vert \Psi ^{v}_{\textbf{k}}(n^{v})\right\rangle$\cite{ACSNano4;1465,CARBON69;151}.
The spectral function, in the presence of a magnetic field, relies on the behaviors of Landau subenvelope functions, which play a important role in determining absorption intensities, excitation channels, and optical selection rules.
This model has been extensively employed to investigate magneto-electronic and magneto-optical properties in few-layer graphenes with uniform stacking\cite{ACSNano4;1465,CARBON69;151,PRB87;205404,SciRep9;859}, findings that have been substantiated by optical experiments\cite{PRL102;166401,PRL100;087403}.
Applying this robust model to stacking modulation in BLG provides a clearer exploration of key aspects, including the energy dispersions of diverse LSs and the optical absorption spectra of this system. These investigations illuminate intriguing phenomena arising from the intricate interactions among electronic states, magnetic confinement, and DW confinement under varying conditions.

\section{Significant Landau subbands induced by domain walls}

In the absence of DWs and external fields, the band structure of AB-stacked BLG features two pairs of parabolic energy bands, primarily contributed by the ${2p_z}$ orbitals within a primitive unit cell\cite{RPP76;056503}. Close to the $K$/$K^{\prime}$ points, one pair of parabolic bands, forming a valence-conduction band crossing at $E_{F}=0$, characterizes it as a zero-gap semimetal. Simultaneously, the second pair experiences an energy shift of $\pm\gamma_{1}$ away from $E_{F}=0$, with states at the K point, denoted as $E^{2^{c,v}}(K)$, representing the band edges.\cite{RPP76;056503}. The first pair is chosen as the focus of study in this work with $N_{B}=50$ modulated with DWs of $N_{D}$. The presence of DWs can profoundly influence the pristine low-lying electronic structures, which undergo significant modulation regions where the stacking configuration experiences periodic variations.

As depicted in Fig. 1(b), DWs with $N_{D}=10$ induce several distorted parabolic bands with split band edges around $k_y a=2\pi/3$. There is a valence-conduction band overlap of 0.03 eV near $E_{F}=0$. The appearance of zone-folding effects around $k_y a=2\pi/3$ occurs at a specific folded $K$ valley due to translational invariance along $\widehat{x}$.
These DWs introduce non-uniformity in the C-C bond lengths, creating a distinct environment that disrupts most symmetries related to $\pm k_{x}$ of pristine AB BLG. In Fig. 1(c), with $N_{D}=20$, the asymmetry in the energy spectrum is enhanced, accompanied by larger band overlap and state splitting within DWs. These changes lead to pronounced alterations in band structures, causing deviations in energy dispersions, band-edge states, and subband hybridizations. As $N_{D}$ further increases, the accumulated DW states near the Fermi level are expected to cause specific magneto-properties. The DW changes introduce unique features such as oscillating and crossing behaviors in specific energy subbands, contributing to the overall complexity of the electronic landscape. Moreover, the Bloch wave functions of low-lying states transition from uniform exponential function to bonding and antibonding standing-wave patterns within DWs.
With increasing subband indices, there is a demonstration of strong hybridizations between neighboring 1D subbands induced by evolving spatial characteristics due to stacking variations\cite{SciRep9;859}. These wave function characteristics redefine the energy subband indices and play a crucial role in shaping the electronic properties of bilayer graphene in the presence of DWs \cite{SciRep9;859}.

In a uniform AB-stacked BLG, well-behaved LLs (gray color) are formed at $\mathbf{B}=B \widehat{z}$ within the first Brillouin zone, as depicted in Fig. 2. The cyclotron centers of LLs, denoted as $d_{c}(k_{y})$ for the $K$ and $K^{\prime}$ valleys, are described as $d_{c}(k_{y})= -k_{y}l_{B}^{2}/4\pi^{2}+ l_{B}^{2}/3\pi a$ and $d_{c}(k_{y})= - k_{y}l_{B}^{2}/4\pi^{2}+ l_{B}^{2}/6\pi a$, respectively \cite{ACSNano4;1465}. The LLs in the two valleys are degenerate and defined as $\alpha$ and $\beta$ states in this study. These LL sequence exhibits a normal energy order characterized by quantum numbers $n^v$=1, 2, 3 and $n^c$=0, 2, 3, and so on, with increasing levels. We anticipate observing variations in the electronic states induced by evolving spatial characteristics due to stacking variations. This phenomenon will be a focal point of the subsequent discussion.

Stacking modulation in BLG introduces diverse Landau quantization effects, dramatically altering the magneto-electronic band structure with the presence of DWs. It significantly impacts the spectral organization of the $\alpha$- and $\beta$-states near the DWs in the supercell. As $l_{B}$ exceeds the DW width, dispersive LSs are formed, exhibiting substantial oscillating amplitudes. For instance, let's consider the (50,10) modulated BLG under $B=65$ T, as depicted in Fig. 2(a). The 2D LL spectrum evolves into a periodic quasi-1D LS spectrum, composed of $\alpha$-states (blue) and $\beta$-states (red). They exhibit symmetry with respect to a phase shift determined by the position of the DWs. Despite the reduction in LL degeneracy, low-lying LSs, particularly $n^{c,v}=0-3$, still exhibit flat dispersions, a characteristic of 2D QLLs localized in Bernal BLG. As $k_{y}$ deviates, energy distortion occurs due to the proximity of $d_{c}$ within the DWs. Moreover, interference between neighboring DWs can induce anticrossing behavior among $\alpha$- or $\beta$-LSs across the energy spectrum.

The influence of DWs on the 1D LS spectrum is accentuated as the magnetic field strength decreases. In Fig. 2(b), the LS spectra of (50,10) BLG at $B=20$ T exhibit significant variations in energy dispersion corresponding to different DW arrangements, contrasting with the relatively flat band structure observed in Fig. 2(a). It reflects that the reduced magnetic quantization accommodates fewer quantum states within the Bernal region. The oscillatory period of LSs increases, and distinct patterns of crossings and anticrossings become more prominent at lower field strengths. As $l_B$ increasingly exceeds the AB-DW boundary at $B=10$ T, the LSs become less well-defined, as depicted in Fig. 2(c), which does not exhibit a flat distribution in the energy spectrum. The presence of QLLs, e.g., $n^{c}=0$ and $n^{v}=1$, which are substantial under magnetic quantization, disappears in this scenario. This suggests that the Bernal width cannot accommodate the full cyclotron motion of the Landau states. Instead, the energy dispersion takes on a parabolic profile due to the confinement of the DW potentials. The low-lying energy bands reflect zone-folding effects, while the magnetic field primarily flattens the energy dispersions near the band-edge states. This transition highlights the intricate interplay between non-uniform magnetic quantization and structural features in determining the electronic properties of the stacking-modulated BLG.

Wider DWs can severely disrupt the flat bands and introduce edge localized states near the Bernal/DW boundary. This is indicative of a higher degree of degeneracy breaking associated with larger stacking modulation. The change in LSs at $B=$ 20 T is more significant in (50,20) BLG, as shown in Fig. 2(d), compared to (50,10) BLG. The former exhibits enhanced LS oscillation and LS anticrossings starting from the Fermi level. However, as exemplified in (50,50) BLG in Fig 2(e), when $N_{D}$ is increased to satisfy the condition $3N_{D}b>l_B$, it implies that QLLs can be stably generated in both the Bernal and DW regions, but the entire k-space is not completely quantized. There exists an uneven energy distribution of the low-lying LSs. The $n^v=0$ and $n^c=1$ LSs disperse across $E_{F}=0$ and extend to deep conduction and valence states. This behavior can be interpreted as the transformation of the magnetically bound states in AB and DW regions. The higher-order subbands, e.g., $n^c,v=2$ and 3 in particular, recover with quantization, while their energies are immediately influenced by the local stacking configuration.

When $B$ increases to 65 T, as expected, the (50,50) system reveals structural differences between DW and Bernal regions concerning $k_{y}$, as demonstrated in Fig. 2(f). Each conduction (valence) LS shows similar dispersion patterns due to the much smaller $l_{B}$. Both types of quantization, occurring in Bernal and DW regions, are regarded as well-defined subbands. A clear demarcation between these two forms of quantization aligns with the AB/DW boundary in the energy spectrum. In this vicinity, individual electron cyclotron motions are permitted across the entire $k$-space, which may reduce LS anticrossing and also benefit the vertical optical excitations.

Based on the aforementioned results, it is evident that the DWs modify the hopping interactions between carbon atoms and alter the energy dispersion. The sensitivity of LS spectra to the structural intricacies introduced by DWs in BLG configurations is emphasized. As either field strength decreases or DW width increases, the suppression of the pristine QLLs intensifies. These 1D oscillating LSs are governed by the periodic arrangement of AB/DW/BA/DW domains, illustrating their dependence on different cases. Under strong magnetic quantization, low-lying flat bands are established. As energy increases, the energy distortion in $k_{y}$ increases rapidly, leading to the occurrence of anticrossing phenomena between $\alpha$ states or $\beta$ states. The interface of LSs results in a broader range of overlapping electron states. These intriguing properties are expected to have a substantial impact on the material's optical excitations.

Meanwhile, LS anticrossings, arising from DW interference, become more prevalent with an increase in the number of DW periods and DW width. The breaking of translational symmetry allows for effective modulation of subband anticrossing in graphene systems, which has been shown to be influenced by factors such as lateral edges in graphene nanoribbons\cite{PCCP18;7573}, and magnetic field\cite{PRB90;205434,PRB83;165443} and electric field\cite{RSCAdv5;80410} in uniform graphene systems. This phenomenon can also be affected by various intrinsic parameters, including interlayer interactions\cite{PRB90;205434} and spin-orbit couplings\cite{PRB94;045410}. The effects of periodic DWs are elucidated in real space in the following sections. However, since $\alpha$ and $\beta$ states correspond to different blocks of DW/AB/DW (DW/BA/DW) in the supercell, the symmetries of energy band, wave function, and optical spectrum are preserved by the two symmetric blocks in the system.

\section{ Density of states}

The DOS serves as a primary reflection of the energy spectra characteristics. Pristine BLG's DOS comprises prominent delta-like peaks, with uniform peak intensity proportional to the magnetic field strength, as observed in the gray curves in Fig. 3. In stacking-modulated BLG, while the such peaks are preserved under strong magnetic quantization, there are several vHSs that reveal the quasi-1D electronic characteristics, including finite flat $k_y$-dispersions, oscillatory dispersions, DW-localized states, and zone-folding effects.

In Fig. 3(a) for (50,10) BLG at $B=65$ T, the $n^{v}=1,2,3$ and $n^{c}=0,2,3$ QLLs are clearly manifested by the prominent peaks (red circles) in the sufficiently wider AB/BA regions and strong field. Also, the emergence of asymmetric peaks between those peaks reflects the band edges of the oscillatory dispersions. The square-root-shaped divergence results from the parabolic extrema. Decreasing $B$ weakens the magnetic quantization and results in significantly reduced peak intensities. For $B=20$ T in Fig. 3(b), the former peaks become faint, and except for the lowest four peaks closest to $E_{F}=0$ that can be clearly defined, the remaining parts exhibit multiple configurations due to DW interference. The slight DOS oscillation below -0.05 eV indicates the LS anticrossings, which are expected to introduce additional selection rules and optical excitations in the system. Conversely, for $B=10$ T (Fig. 3(c)), all the DOS peaks exhibit 1D-square-root divergence, in which the vHSs come from the parabolic band edges of localized states in DWs. This suggests that the dominant quantum confinement gradually supersedes the minor magnetic field effects as $l_{B}$ approaches the system boundary.

Adjusting the value of $N_{D}$ brings about more changes in the quasi-1D peaks related to the DWs. Figure 3(d) reveals an intriguing transformation of the vHSs in (50,20) BLG at $B=20$ T as DW-interference enhancement with wider DWs. The vHSs from QLLs correspond to those of (50,10) in Fig. 3(b), while more vHSs are induced by the DW-localized states amplified with the LS anticrossings. In the case of systems such as (50,50) in Fig. 3(e), these effects become more pronounced with longer stacking modulation. Enhanced vHSs are observed, accompanied by blue shifts of QLL peaks from both regions, which coexist and predominate in the low-energy DOS (red symbols). The recovery of $n^{c,v}=3$ peaks reflects the unconventional QLLs associated with DW regions. The modulation-induced changes in DOSs are more readily apparent under these conditions. Moreover, the peak originating from the two lowest LSs is of particular importance and determines the threshold channel in the optical spectrum.

However, at $B=65$ T, each well-quantized LS contributes to a symmetric peak and two asymmetric peaks (Fig. 3(f)). Such grouped peaks are well separated, indicating dominating optical excitations due to the obvious joint DOS. Overall, these pronounced changes in the aforementioned different cases directly reflect the effects of the magnetic field on stacking-modulated BLG. These alterations in the DOS and the emergence of vHSs are crucial factors to consider when understanding the magneto-optical properties in the stacking-modulated BLG.

\section{ Magnetic wave functions}

In pristine BLG, the dependence of Landau wave functions on $k_{y}$ and $n^{c,v}$ has been previously explored in the supercell\cite{ACSNano4;1465}. Research on few-layer graphene has identified a distinct LL sequence depending on stacking symmetries\cite{PRB90;205434}. However, the reduced magnetic confinement on electrons enables them to spread out over a larger area in graphene sheets and interact more with neighboring DWs. This section illustrates the evolution of the wave functions for the two kinds of sublattices A and B in the supercell, which contains $N_{sc}I_{sc}$ atoms for each kind of sublattices in each graphene sheet, where $I_{sc}=2(N_{B}+N_{D})$. Here, we focus on $\alpha$-states, which are symmetric with $\beta$-states.

In the case of (50, 10) BLG, the representative states, labeled $\alpha_{1}$-$\alpha_{6}$, illustrate the wave function evolution at $B=65$ T, as shown in Fig. 4. The $K$ valley localized states are distributed around 2/3 of the enlarged supercell\cite{ACSNano4;1465}. $\alpha_{1}$ to $\alpha_{3}$ represent well-defined Landau states in the Bernal stacking, maintaining predominantly symmetric relationships between the four inequivalent sublattices. Specifically, these states exhibit a quantum mode difference between the sublattices, with values of 1 and 0 under the connection of $\gamma_{0}$ and $\gamma_{1}$, respectively\cite{ACSNano4;1465}. The dominating sublattice $B^1$, used to identify the quantum numbers of LLs, retains a symmetric distribution due to its separation from the boundary. Interfered in this context, the left Bernal stacking refers to the BA configuration.

On the other hand, the wave functions display strong $k_y$ dependence, reflecting the magnetic localization center gradually approaching, entering, and eventually crossing the left DW. From $\alpha_{4}$ to $\alpha_{6}$ in the $n_{v}=2$ LS, the wave functions gradually penetrate into the DW (black dot area). In the vicinity, both the symmetry and sublattice quantum modes gradually break down. After the DW, the stacked structure undergoes a transformation from BA to AB, resulting in a correspondence change from ($A^1$, $B^1$, $A^2$, $B^1$) in the former region to ($B^2$, $A^2$, $B^1$, $A^2$) in the latter region. As a result, the emergence of the quantum number $n_{v}=2$ is reinstated by $A^2$ for $\alpha_{6}$ state in AB configuration, which clearly indicates a change in the dominating sublattice influenced by the stacking modulation in the graphene system.

Based on the behavior of the wave functions, the LSs in the entire $k$ space can be classified as localized states in either Bernal or DW-influenced regions, as indicated, respectively, by green and purple hollow dots in Fig. 4(a). In each $\alpha$ LS, flat parts corresponding to different Bernal regions are separated by a DW in the supercell. It explains that as $n^{c}$ increases, the QLLs become narrower, and DW interference becomes more prominent. When the influence of DWs exceeds the spacing between neighboring LSs, frequent LS anticrossings become apparent in the magneto-electronic spectra. These anticrossing phenomena are characterized by significant hybridization of magnetic subenvelope functions, denoted as major and minor modes\cite{PRB90;205434, RSCAdv5;80410}. These effects are expected to impact the optical absorptions.

To illustrate the special magnetic quantization within DWs, we examine the (50,50) system at $B=20$ T in Figs. 5(a)-5(c).
As for the lowest $n^{c}=0$ LS, the dominating sublattice $A_{2}$ is revealed by $\alpha_{1}$ state, which is predominantly localized in the AB region (Fig. 5(b)). As $k_{y}$ moves away, the zero-mode symmetric wave functions are preserved for $\alpha_{2}$ and $\alpha_{3}$ in a nearby DW while showing different subenvelope relationships, where the identical mode appears for $B^{1}$ and $A^{2}$ ($A^{1}$ and $B^{2}$). For higher states, wider wave function distribution across the AB/DW boundary follows the identifiable feature. In particular, the state hybridization associated with two neighboring DWs is demonstrated by the anticrossing states, e.g., $\alpha_{4}$, $\alpha_{5}$ and $\alpha_{6}$ in Fig. 5(c). Such states formed near the boundary clearly separate the spectral features of the quantized states in the band structure. It is anticipated that optical excitation can be enriched and diversified by the interplay between two kinds of QLLs.

When $l_B$ is significantly smaller than the DW width, e,g, $B=65$ T, we discuss the well-defined Landau quantization in both Bernal and DW regions for larger energies by examining several conduction LSs, as shown in Fig. 5(d). For $n^{c}=0$, the transition of localization from Bernal to DW regions is illustrated by $\alpha_{1}$ to $\alpha_{3}$ in Fig. 5(e). This transition is characterized by the dominating sublattice, alternating the shift from Bernal to DW. As the strong field confines the cyclotron motion spanning across multiple DWs, the behavior is preserved up to higher LSs, such as $\alpha_{4}-\alpha_{6}$ corresponding to $n^{c}=3-5$. Moreover, the boundary between DW and Bernal regions is clearly observed at a fixed $k$ value throughout the entire $k$ space.

Several aspects have been observed in the subsequent changes of the above wave functions, including the varying degree of the DW barrier, the substantial transition of dominating sublattices, symmetry breaking, and the mode changes on these sublattices. It should be noted that DWs disrupt the hexagonal symmetry and AB symmetry in the system. The anticrossing patterns manifest non-trivial quantum number differences across stacking boundaries. This indicates that the effective-mass approximation is not suitable for accurately determining the unconventional magnetic subenvelope functions under such non-uniform intralayer and interlayer atomic interactions, as each Landau state may consist of a superposition of various modes.

\section{ Magneto-optical properties}

The magneto-optical properties of graphene is profoundly affected by the presence of DWs in BLG.
In Fig. 7, we investigate the effects of DWs on the (50,10) BLG system at decreasing $B$ from $65$ T to $10$ T.
The unique absorption features vary with different DW barriers under nonuniform magnetic quantization.
At $B=65$ T, as shown in Fig. 7(a), the pristine magneto-absorption spectrum, denoted by the gray curve, is characterized by vertical transitions between highly-degenerate LLs\cite{ACSNano4;1465}. These peaks, non-equally spaced with delta-function divergence, satisfy the selection rule $\Delta n=n^{c}-n^{v}=\pm1$, while exhibiting approximately uniform intensities due to the equal degeneracies of each LL. For instance, the first three peaks, marked by diamond symbols, correspond to inter-LL transitions between ($n^{v},n^{c}$)= (1,2), (3,2), and (2,3), while the fourth peak arises from symmetric channels (3,4) and (4,3).

The stacking-modulated absorption features are depicted by the blue curve in Fig. 7(a). The optical excitations occur exclusively between spatially separated $\alpha$ or $\beta$ states, which contribute equally to the material's absorption spectrum. These optical resonances exhibit intensity discrepancies throughout the system, leading to an increased number of peaks due to the emergence of more localized states. However, the optical excitations between the QLLs and oscillating subbands give rise to unique spectral structures. The former, preserved under strong magnetic quantization, remains as prominent delta-function peaks traced from the same inter-LL transition frequencies, albeit suppressed by the contraction of the flat bands. The latter results in the emergence of asymmetric square-root peaks with relatively weak intensity, associated with the vHSs at the band edges. In the insert, these two kinds of channels are illustrated by the band structure. Indeed, despite the dominance of the pristine selection rule $\Delta n=\pm1$, an additional $\Delta n=\pm2$ is observed for the second peak. Meanwhile, peaks characterized by $\Delta n=0$ and $\Delta n=\pm2$ are also evident for those DW-localized states. As the magnetic field confinement diminishes within the DW regions, optical excitations coincide with the enrichment of the localized states.

Reducing $B$ to 20 T in Fig. 7(b), a red shift and weakened absorption intensity are observed in the (50,10) system. The pristine peaks of $\Delta n=\pm1$ are absent because the flat bands obscure for $n^{c,v}\geq 2$. However, the evidence of $n^{v}=0$ QLL is presented by the threshold peak at $\omega\simeq0.03$ eV. The spectral features transition to the predominance of various DW-induced quasi-1D vHSs, following different selection rules $\Delta n=0$, $\Delta n=\pm1$, and $\Delta n=\pm2$. This expansion occurs because the influence of DWs extends beyond the cyclotron motion of electrons in the system. Furthermore, the absorption characteristics vary with the DW interference, especially in the LS anticrossing region, where excitation enhancement is evident through the multiple peaks (green symbols), with the most apparent peak observed at $\omega \gtrsim 0.15$ eV. This behavior is consistent with the anticrossing bands for $E^{v}\lesssim -0.05$ eV and $E^{c}\gtrsim 0.07$ eV in Fig. 2(b). In such regions, the state hybridization leads to the emergence of multiple selection rules and a significant enhancement in both the sharpness and intensity of absorption peaks.

At $B=10$ T in the (50,10) system, Fig. 7(c) illustrates a further broadened spectrum due to increased overlapping electronic states within the narrower LSs. In contrast to the previous cases, the spectral features does not follow the typical behavior associated with QLLs, and multiple selection rules dominate across the entire excitation spectrum. The optical transitions are permitted between every valence and conduction LSs as the whole spectrum evolves into parabolic dispersions, evident from the square-root DOS peaks in Fig. 3(c). The quantum confinement effect competes with magnetic fields as $l_{B}$ gradually approaches the system scale, ultimately playing a crucial role in shaping the 1D optical characteristics\cite{SciRep9;859}. Furthermore, the threshold peak, regarded as indicative of the lowest $n^{c}=0$ and $n^{v}=1$, remains prominent but experiences a significant red shift. In such a scenario, it becomes challenging to distinguish these two types of absorption channels as $l_{B}$ substantially extends beyond the Bernal region. This reflects the fact that the quantization phenomena become less well-defined because electrons gain more freedom in their energy states.

In Fig. 8, we examine the impact of DW size on the magneto-absorption spectrum, in contrast to the quantization phenomena observed in Fig. 7 with respect to $l_{B}$. Increasing $N_{B}$ induces notable changes, with larger DWs exhibiting oscillating LSs of greater amplitudes and thus a broader spectrum compared to smaller DWs at the same field strength, as observed in (50,10) and (50,20) at $B=20$ T (Figs. 7(b) and 8(a)). The blue curve in Fig. 8(a) illustrates disrupted original peaks and newly emerged peaks that deviate from typical behavior, attributed to several subchannels from wider DWs. Stacking modulation becomes evident in the LS anticrossing region, where some peaks emerge in the spectrum for $\omega \gtrsim 0.15$ eV, due to significant LS breaking of the hybridized valence and conduction LSs.

Conversely, when $l_B$ is smaller than the DW width, as illustrated by the (50,50) BLG at $B=20$ T (red curve in Fig. 8(a)), a different scenario unfolds. The center of QLLs remains stable within individual Bernal regions, DWs, or near the Bernal/DW boundaries, which facilitates inter-band transitions from those unconventional QLLs. As a result, the spectra reflect the structural differences between the DW and the Bernal regions. The initial channel $n^{v}=1\longleftrightarrow n^{c}=0$ governs the first few peaks, extending the energy range of $0\leq\omega\lesssim 0.1$ eV due to stacking modulation. Meanwhile, the emergence of large-scale DWs triggers a prominent composite peak at $\omega\simeq0.16$ eV, as contributed by excitations from magnetically quantized states stably formed in the both regions.

At a stronger magnetic field strength of $B=65$ T, the (50,50) system, depicted in Fig. 8(b), reveals prominent magneto-spectral structures. Some spectral peaks coincide with those in pristine BLG, appearing around ($n^{v},n^{c}$)= (1,2), (3,2), (2,3), and (3,4). However, other asymmetric peaks are associated with inter-QLL transitions within DWs. These transitions follow various selection rules, such as $\Delta n=0$, $\Delta n=\pm1$, and $\Delta n=\pm2$. Such induced channels become gradually more considerable across the specific excitation spectrum, exemplified by the strong peak highlighted at $\omega\simeq0.25$ eV, which is absent in the (50,10) system. In short, the varied cases with different $l_{B}$ and $N_{D}$ contribute to enriched and diversified absorption spectra, providing valuable insights into the system's optical response under stacking modulations.

Experimental verifications through STM and various optical spectroscopies have provided valuable insights into the electronic quantization and optical responses in low-dimensional systems. For instance, STM measurements on
finite-length carbon nanotubes have identified well-behaved standing waves with specific zero points\cite{Science283;52}. Similar measurements on graphene nanoribbons have revealed the edge states\cite{NatPhys7;616,NatComm13;7814} and scattering standing waves\cite{NanoLett11;3663}, demonstrating the interplay between finite-size confinement, edge structures, and magnetic fields. Landau quantization of topological edge states created by the AB-BA DW in BLG has also been identified\cite{NatComm7;11760}, displaying distinct features in the 2D LL energy spectra observed in few-layer graphene\cite{PRB91;115405}.
Besides, in Bernal graphite, 3D magnetically quantized states display distinct spectral characteristics associated with the K and H points\cite{NatPhys3;623}. Magneto-optical spectroscopy has also confirmed the optical transitions in graphite, presenting their significance in the context of Bernal graphite and Bernal BLG\cite{PRL102;166401}.

Compared to these systems, it is important to verify the theoretical calculations in the stacking-modulated BLG, particularly focusing on the distinct oscillating LSs and absorption spectra, as well as the spatial magnetic wave functions in the supercell. These DWs enhance the confinement of electronic states, weaken magnetic quantization, create unconventional QLLs and trigger extra absorption peaks. Moreover, verifications should be taken on anticrossing effects that influence the intricate electronic and spectroscopic features, which reveals the interplay between DWs and external fields in low-dimensional systems.

\section{Concluding remarks}

In conclusion, this study extensively explores the unique electronic and optical properties of stacking-modulated BLG under external magnetic fields. Stacking modulation dramatically alters the Landau quantization effects, reshaping the magneto-electronic band structure contributed by ${2p_z}$ orbitals within a supercell. The introduction of periodic DWs profoundly influences significant modulation regions where the BLG configuration experiences stacking variations. We analyze localized atomic wave functions and corresponding lattice symmetries within a generalized tight-binding model. The presence of DWs disrupts translational symmetry, transforming 2D QLLs into two types of quasi-1D localized states, referred to as $\alpha$- and $\beta$-LSs. The observed 1D features in stacking-modulated BLG are unique compared to the 2D phenomena observed in sliding\cite{SciRep4;7509} and twisted\cite{PRB87;205404} BLG. Additionally, the differences in LSs are sharply contrasted with graphene nanoribbons\cite{PCCP18;7573} and graphite\cite{Carbon43;1424}.

Varying DWs manifest distinctive quasi-1D dispersions related to length scales between the magnetic fields and the system. The unique LS features, such as flat and oscillating subbands, as well as crossing or anticrossing behaviors at specific energies, are introduced. These modifications, particularly evident in their LS energies, are more pronounced for smaller $N_{D}$ and larger $l_{B}$, indicating the breaking of LLs due to the cyclotron motion of electrons affected by the DW barrier and the stacking boundary. Additionally, the magnetic wave functions also undergo significant variations due to DW interactions, responsible for the distinctive quantization patterns and hybridization phenomena. Furthermore, the DOS reflects the energy spectra characteristics, unveiling quasi-1D electronic features induced by non-uniform quantization and DW interference.

The absorption spectral features align with magnetoelectric properties, with optical excitation efficiently modulated between spatially separated $\alpha$ or $\beta$ states. The pristine inter-LL transitions are permitted for flat bands, possibly formed in the Bernal region, but they are distorted by the DW barrier, inducing additional selection rules. The observed LS anticrossings are a significant outcome of DWs in BLG, affecting a broad spectrum of absorption features. Structural disparities between DW and Bernal regions yield distinct magnetically quantized states, impacting interplay between LSs and the absorption spectrum. Remarkably, the threshold peak serves as evidence for the emergence of the lowest QLL and remains robust even in the presence of DWs in BLG. These essential properties align with the alternating dominance between magnetic and quantum confinement effects on the spectrum.

Understanding the behaviors of quasi-1D LSs is crucial for comprehending stacking-modulated BLG properties. Experimental verifications are essential to validate theoretical calculations, focusing on distinct oscillating LSs, absorption spectra, and spatial magnetic wave functions in stacking-modulated BLG. The interplay between DWs, magnetic fields, and structural features significantly influences BLG's electronic and optical properties. These features are experimentally observable and offer new possibilities in magneto-electronics, optoelectronics, and quantum technologies.

\section{Acknowledgments}
This work was supported in part by the National Science Council of Taiwan, the Republic of China, under Grant Nos. NSC 109-2112-M-006 -023 -MY2. GG acknowledges the support from the US AFRL Grant No. FA9453-21- 1-0046.

\newpage
\renewcommand{\baselinestretch}{0.2}


\newpage
\begin{figure}
\centering \includegraphics[width=0.9\linewidth]{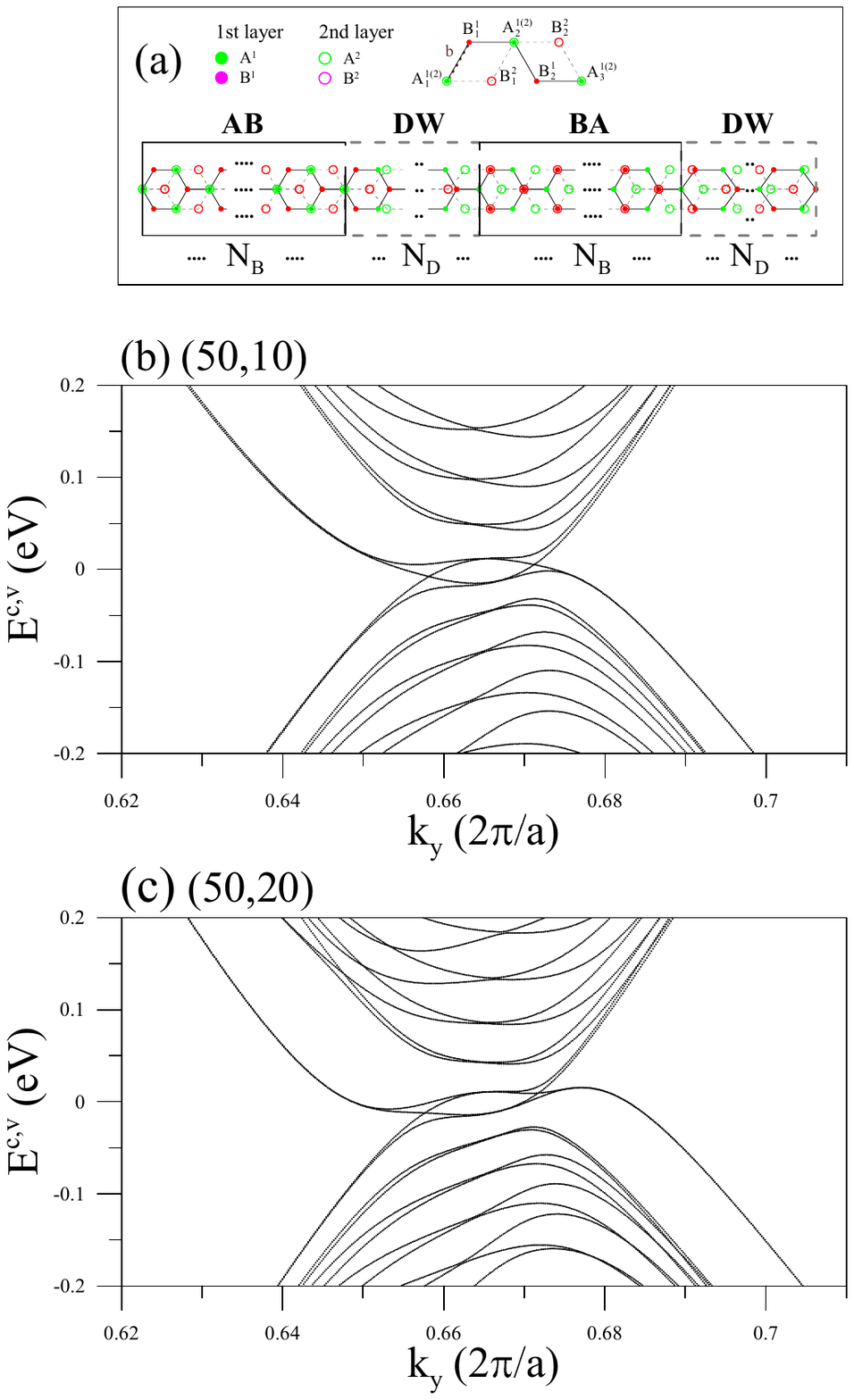}
\begin{center} Figure 1: (a) Unit cell of stacking-modulated BLG with the AB/DW/BA/DW configuration, described by parameters $N_{B}$ and $N_{D}$, where the two Bernal (AB and BA) or DW widths are equal. Sublattice indices are illustrated above. The C-C bond length $b$ is equal to 1.42 ${\AA}$. Energy band structures of (b) ($N_{B}$, $N_{D}$)=(50, 10) and (c) (50, 20) systems as a function of $k_{y}$ in $2\pi/a$ scale, with $a=\sqrt{3}$ $b$ as the graphene lattice constant.
\end{center} \end{figure}

\newpage
\begin{figure}
\centering \includegraphics[width=0.9\linewidth]{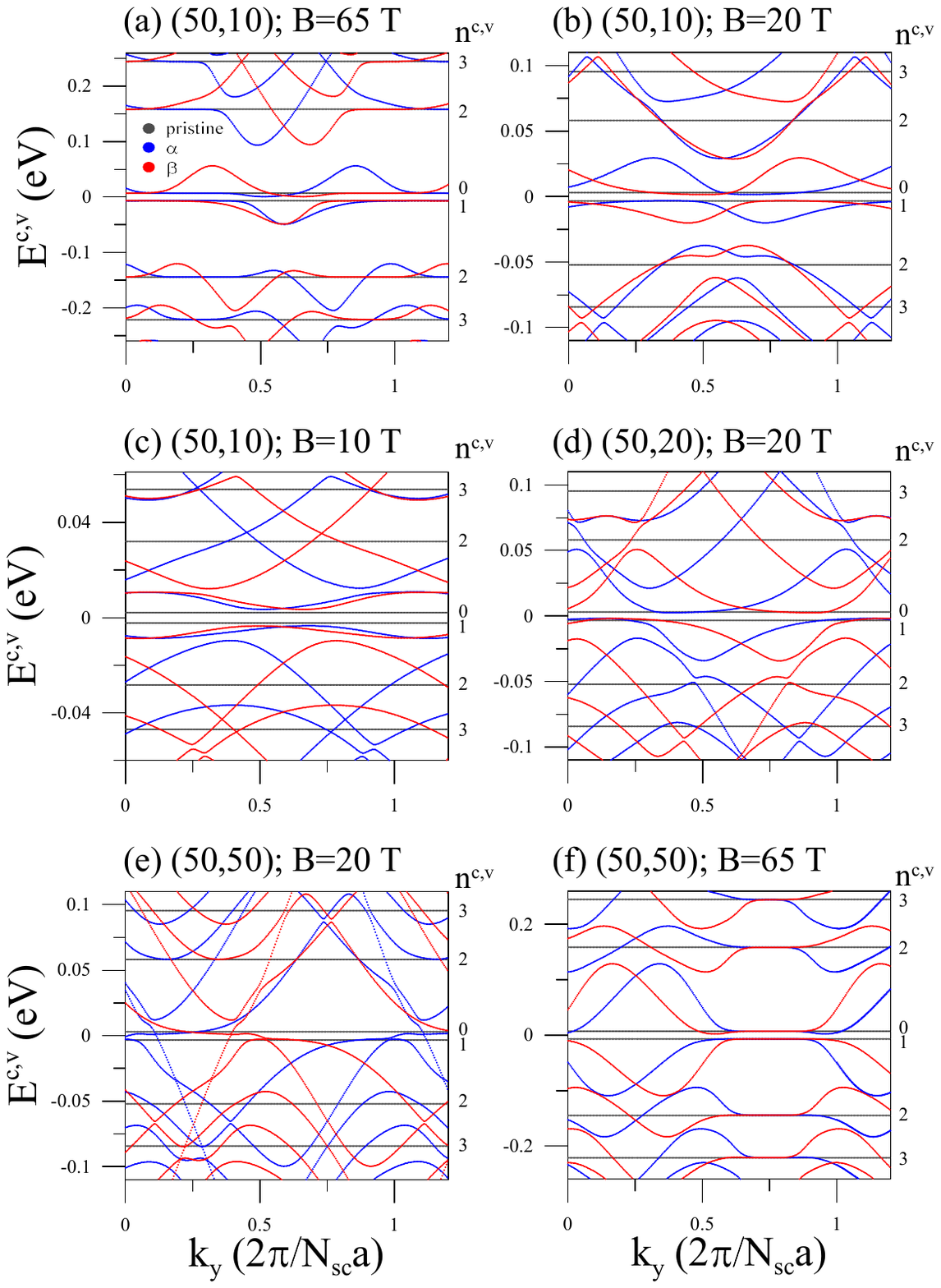}
\begin{center} Figure 2: $\alpha-$ and $\beta-$LS spectra for the (50, 10) system are displayed at (a) $B = 65$ T, (b) $B = 20$ T, and (c) $B = 10$ T; for the (50, 20) system at $B = 20$ T in (d), along with the (50, 50) system at $B = 20$ T in (e). Lastly, (f) illustrates the (50, 50) system at $B = 65$ T. LL energies calculated for pristine BLG are included for comparison. Here, $n^{c,v}$ denotes the conduction and valence subband indices.
\end{center} \end{figure}

\newpage
\begin{figure}
\centering \includegraphics[width=0.9\linewidth]{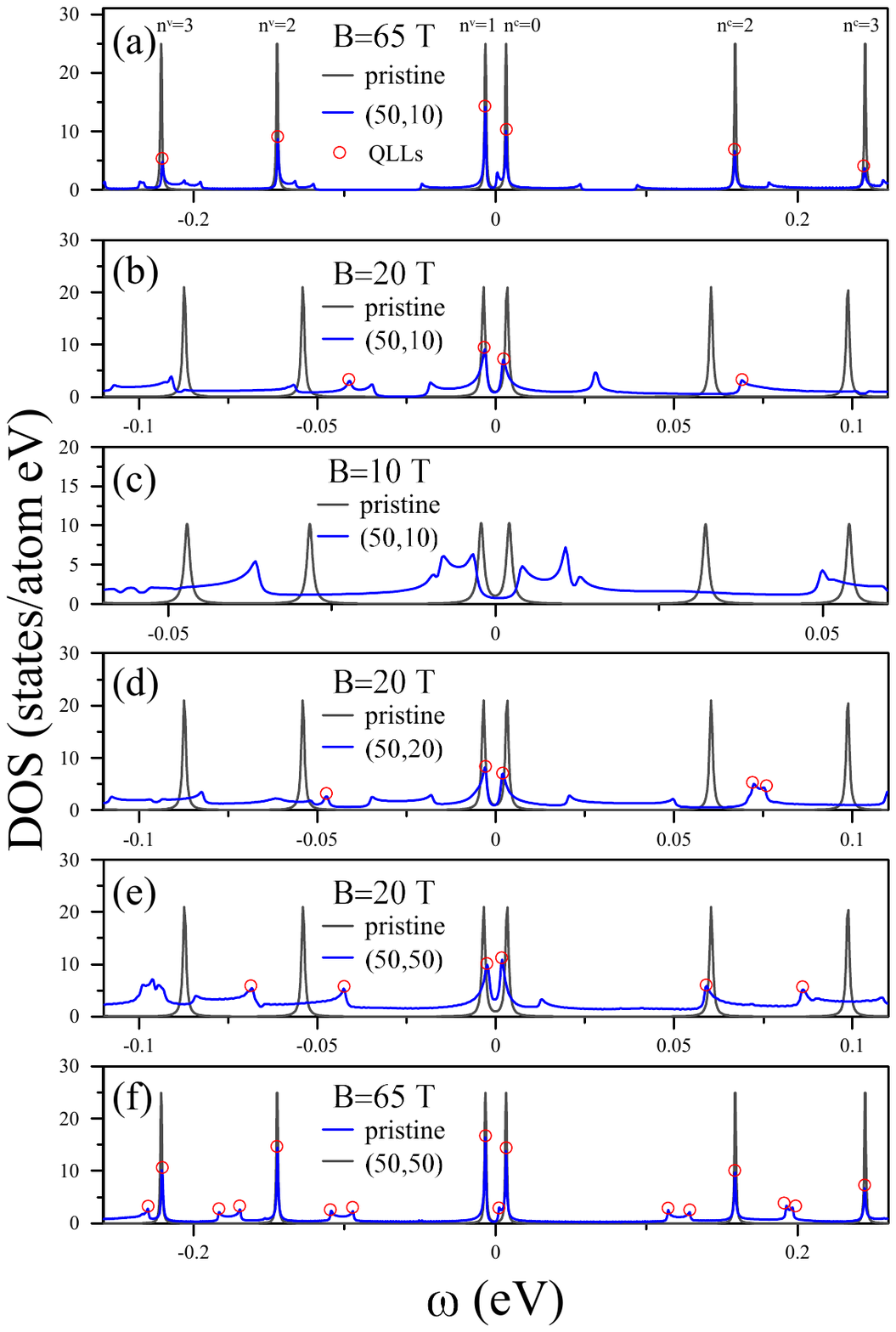}
\begin{center} Figure 3: DOS of LSs for BLG featuring various modulated stacking configurations and magnetic field strengths. Red hollow symbols denote the vHs arising from QLLs.
\end{center} \end{figure}

\newpage
\begin{figure}
\centering \includegraphics[width=0.9\linewidth]{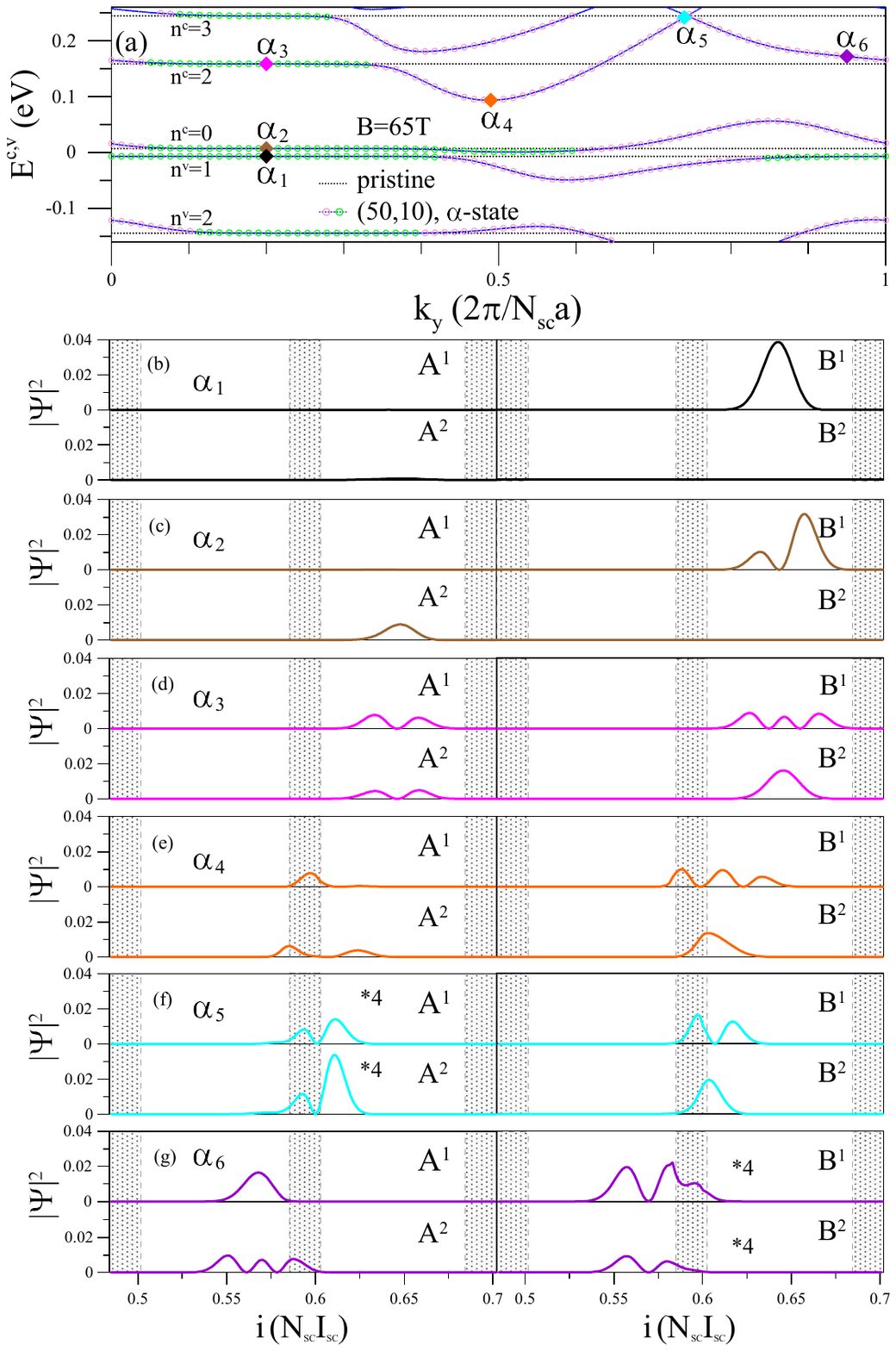}
\begin{center} Figure 4: (a) The first five LSs for the (50,10) system at $B=20$ T, where green and pink colors denote localized states in Bernal and DW regions, respectively. (b)-(g) Subenvelope functions of $A^1$, $B^1$, $A^2$, and $B^2$ for the $\alpha_{1}$-$\alpha_{6}$ states in (a) are plotted as a function of sublattice index $i$ normalized by $N_{sc}I_{sc}$. The black-dotted areas indicate the extents of the DWs.
\end{center} \end{figure}

\newpage
\begin{figure}
\centering \includegraphics[width=0.9\linewidth]{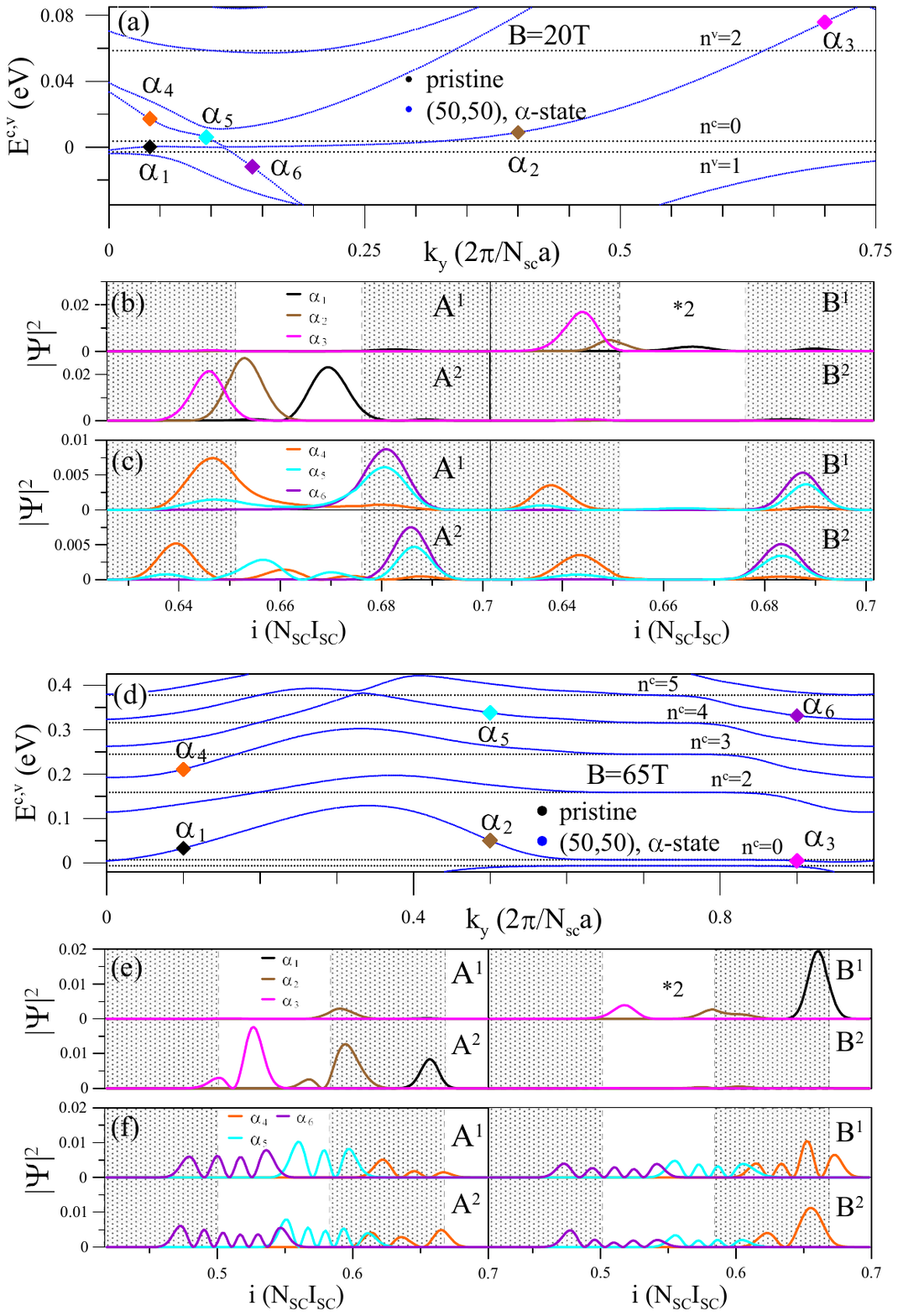}
\begin{center} Figure 5: A plot similar to Fig. 4 for the (50,50) system at $B=20$ T in (a)-(c) and for the (50,50) system at $B=65$ T in (d)-(f).
\end{center} \end{figure}

\newpage
\begin{figure}
\centering \includegraphics[width=0.9\linewidth]{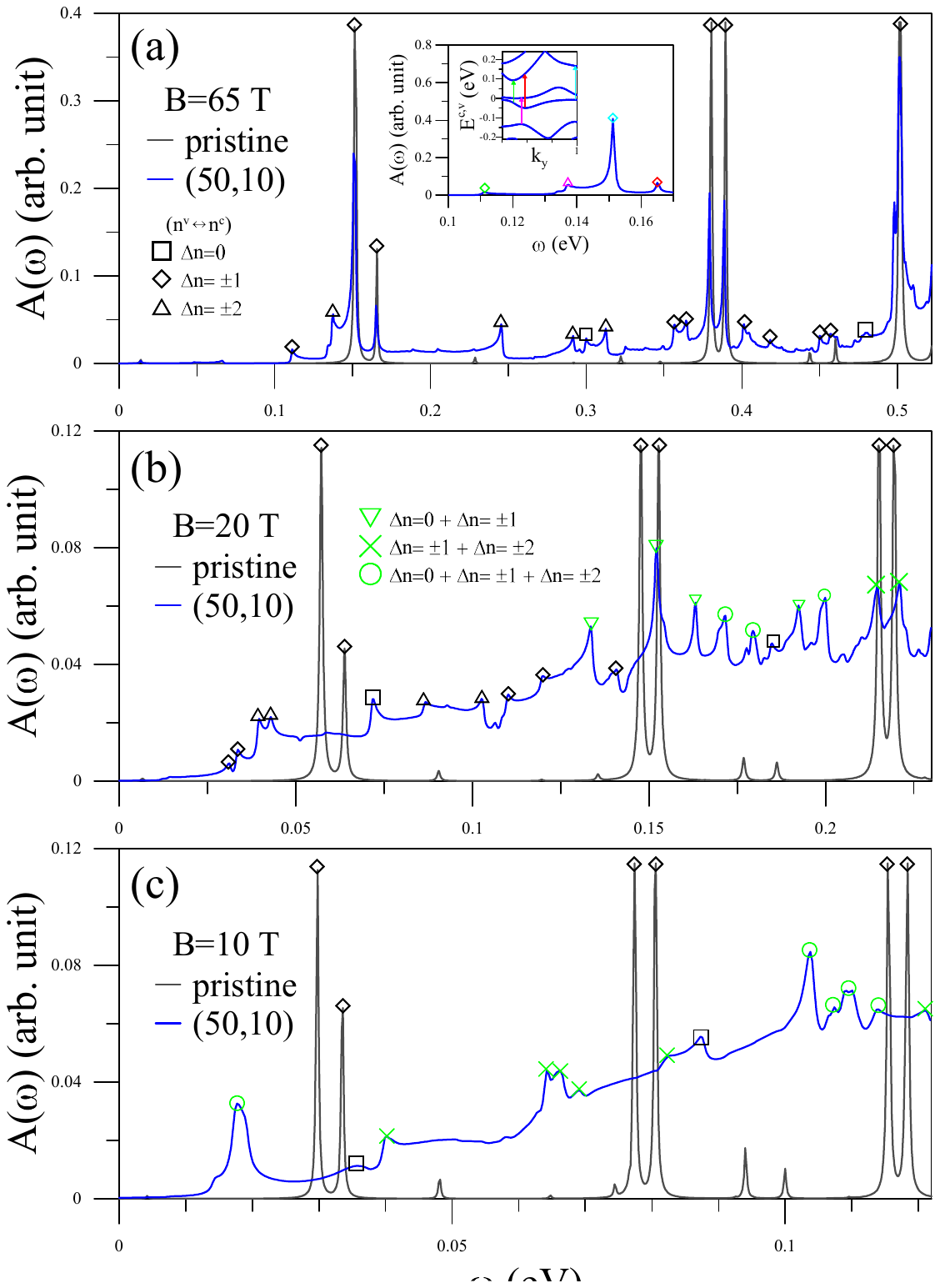}
\begin{center} Figure 6: Magneto-optical absorption spectra are depicted for (50,10) stacking-modulated BLG at (a) $B=65$ T, (b) $B=20$ T, and (c) $B=10$ T. The lowest four channels are illustrated by the LS spectrum in the insert. Hollow symbols indicate the optical selection rule for the absorption peaks. The spectrum conducted in pristine BLG without DWs is shown by gray curves for comparison.
\end{center} \end{figure}

\newpage
\begin{figure}
\centering \includegraphics[width=0.9\linewidth]{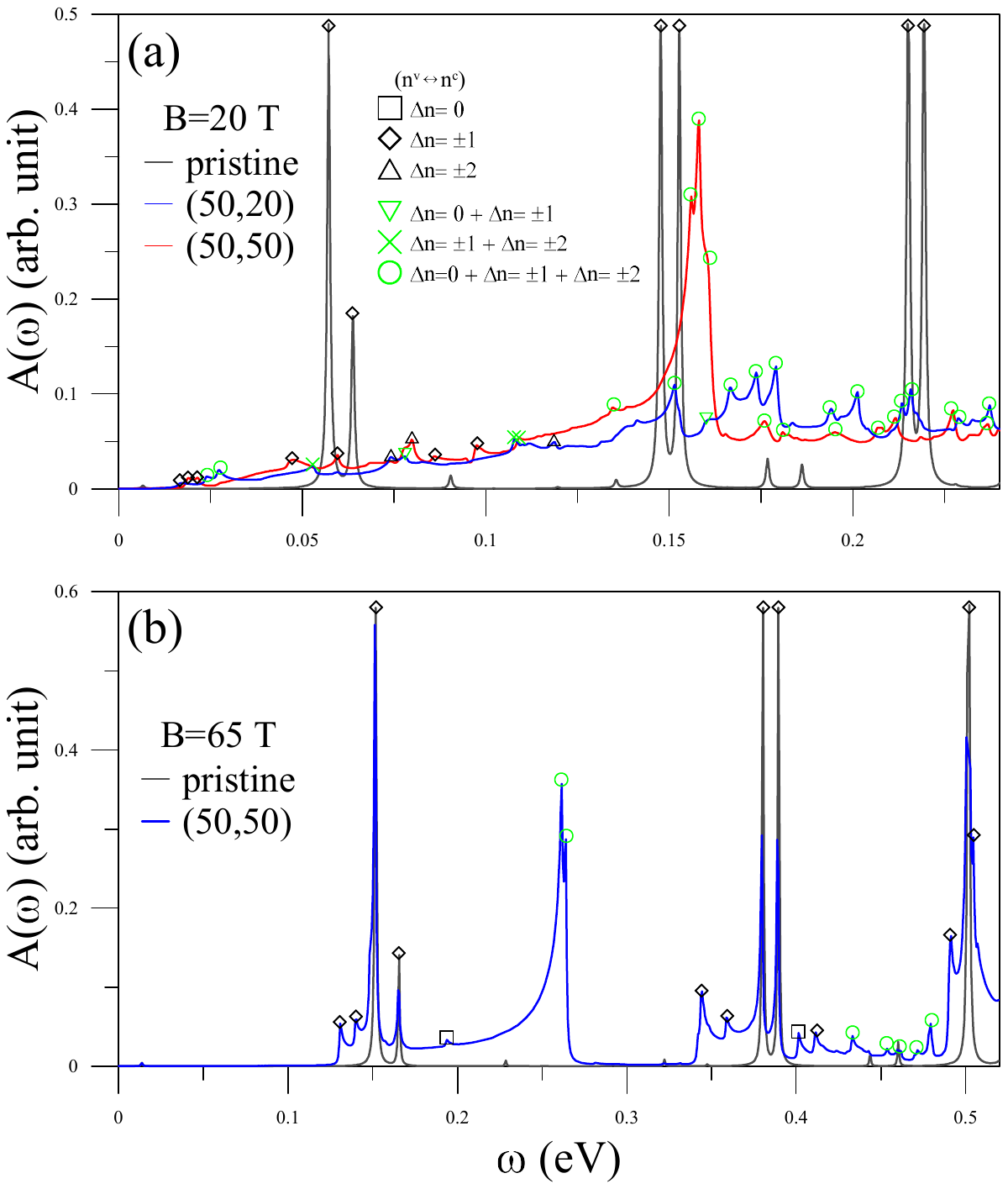}
\begin{center} Figure 7: Magneto-optical absorption spectra for (a) (50,20) and (50,50) stacking-modulated BLGs at $B=20$ T, and (b) (50,50) system $B=65$ T. The spectrum conducted in pristine BLG without DWs is shown by gray curves for comparison. Hollow symbols indicate the optical selection rule for the absorption peaks.
\end{center} \end{figure}

\end{document}